\documentclass[conference]{IEEEtran}
\usepackage{amsmath, amssymb, graphicx, algorithm, algorithmic, cite}
\usepackage[english]{babel}

\usepackage{subfigure, color, stfloats}

\usepackage{flushend}

\begin{document}

\title{Truly Intelligent Reflecting Surface-Aided Secure Communication Using Deep Learning}

\author{\IEEEauthorblockN{Yizhuo Song$^1$,
        Muhammad R. A. Khandaker$^1$,
        Faisal Tariq$^2$, 
        Kai-Kit Wong$^3$, and
        Apriana Toding$^4$
        } 
\IEEEauthorblockA{
        $^1$School of Engineering and Physical Sciences, Heriot-Watt University, Edinburgh, United Kingdom\\
        $^2$James Watt School of Engineering, University of Glasgow, United Kingdom\\
        $^3$Department of Electronic and Electrical Engineering, University College London, United Kingdom\\
        $^4$Dept. Electrical Engineering, Faculty Engineering, Universitas Kristen Indonesia Paulus, South Sulawesi, Indonesia\\
Corresponding e-mail: $\{\rm ys32, m.khandaker\}@hw.ac.uk$}}

\maketitle

\begin{abstract}

This paper considers machine learning for physical layer security design for communication in a challenging wireless environment. The radio environment is assumed to be programmable with the aid of a meta material-based  intelligent reflecting surface (IRS) allowing customisable path loss, multi-path fading and interference effects. In particular, the fine-grained reflections from the IRS elements are exploited to create channel advantage for maximizing the secrecy rate at a legitimate receiver. A deep learning (DL) technique has been developed to tune the reflections of the IRS elements in real-time. Simulation results demonstrate that the DL approach yields comparable performance to the conventional approaches while significantly reducing the computational complexity.

\end{abstract}


%
\IEEEpeerreviewmaketitle

\section{Introduction}

Security has become a major concern for wireless communication systems with the emergence of high data rate and low latency requirements which, due to inherent vulnerability in their architecture, has limited ability to embed security in the higher layers of communication. Thus physical layer security (PLS) will become an integral part of future communication systems beyond 5G \cite{6g_vision}. Despite rapid progress in PLS techniques, the effectiveness of PLS in real scenarios are in doubt due to some major challenges including energy cost incurred in relaying jamming signals and artificial noise as well as computational complexity in beamforming design \cite{yu2019IRS_PLS}.


With the development of metamaterials technology, intelligent reflecting surface (IRS) has emerged as a promising technique for future wireless communications due its ability to reconfigure the wireless propagation environment by exploiting a large number of low-cost passive reflection units (thus incurs no energy cost for reflecting the signals) which can intelligently adjust the incident signal to improve the system performance \cite{wu2019beamforming}.
In particular, IRS has great potential in enhancing physical layer security  \cite{GuanIntelligent} by intelligently tailoring the multipath propagation. By adjusting the phase shift of reflection unit adaptively, the signal reflected by IRS can be enhanced or weakened correspondingly at the receiver, thus strengthening the desired signal and attenuating interference signal \cite{cui2019secure}. The secrecy  rate can be greatly improved by jointly optimizing the beam-forming at the base station and the phase shifts of IRS units. However, generating the optimal phase shifts of IRS elements with acceptable computational complexity remains as the most pressing challenge \cite{wu2019beamforming}.

In \cite{feng2019secure}, a two phase optimisation approach is adopted in which closed form expression for beamforming design followed by identification of appropriate phase shift in IRS reflector. In \cite{shen2019secrecy}, secrecy rate maximisation algorithm was developed and a closed form and semi-closed for expressions were obtained for beamforming and phase shift of IRS respectively. In \cite{yu2019IRS_PLS}, the algorithms of block coordinate descent (BCD) and minorisation maximization (MM) are investigated for small-scale IRSs and large-scale IRSs, respectively. Moreover, the authors of \cite{chen2019intelligent} proposed an iterative path-following algorithm. It was shown in \cite{li2019joint} that a locally optimal solution can be obtained by using second-order cone programming, which achieves better performance than conventional semidefinite programming (SDP) based algorithms. However, all the above iterative algorithms involve heavily computation-demanding matrix inversions, and their  complexity increases exponentially with the number of IRS reflection units. In addition, the alternating optimization algorithms usually require long time to find an apparently optimal solution, which makes the solution less attractive for practical implementation.

Recently, machine learning (ML) and deep learning (DL) techniques have attracted huge interest for addressing wireless communication problems due to their ability to improve system performance and reduce the computational cost \cite{JGao19}. DL exploits the deep neural network (DNN), which completes the training process offline and the trained DNN only includes simple linear and non-linear transformation units.  
Although the name suggests that IRSs are intelligent, existing works with IRS-assisted communications do not implement any systematic learning approach. While ML/DL has been widely investigated in wireless communications, there is hardly any work that exploits DNN for IRS and to the best of the authors' knowledge, there is no such work that exploits ML for IRS-aided PLS. In this work, DL is applied to design the optimal reflection coefficients of the IRS elements for secure communication. The main contributions of the paper include: (i) a deep learning solution for designing truly intelligent reflecting surface for the secrecy rate maximization problem, and (ii) rigorous performance analysis and comparison with existing optimisation algorithms. The proposed DNN approach provides us real-time IRS reflections thus addressing the implementation challenges of iterative IRS design.

\textbf{Notations--} In this paper, bold lower-case and upper-case letters represent vectors and matrices, respectively, $\mathbb{C}^{M \times N}$ denotes the space of $M \times N$ complex-valued matrices, $\mathbf{H}^T$ stands for the transpose of matrix $\mathbf{H}$, while $| \cdot |$ and $\|\cdot\|$ denote absolute value and the Euclidean norm, respectively. $\operatorname{diag}({\bf g})$ means a $N \times N$ diagonal matrix with ${\bf g} \triangleq [g_1, ~g_2,~\cdots, ~g_N]$ as the main diagonal and $[ \cdot ]^+$ denotes $\max(0,x)$.

\section{System Model and Problem Formulation}\label{sec_sys_mod}
\begin{figure}[ht]
\centering
\includegraphics[width=0.8\linewidth]{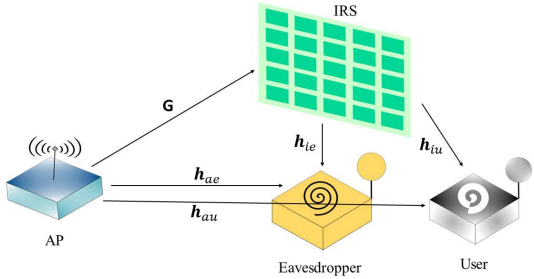}
\protect\caption{An IRS-assisted wireless communication system.} \label{SystemModel}
\end{figure}

Fig.~\ref{SystemModel} illustrates the IRS-assisted wireless system model of interest, which consists of an access point (AP), one legitimate user, one eavesdropper and an IRS. The AP transmits signals to the user in the presence of the eavesdropper. We assume that the AP is equipped with $M$ antennas, while the user and the eavesdropper each with a single antenna. The IRS is deployed in the network between the AP and the user to aid secure data transmission to the user, with $N$ reconfigurable reflecting units programmed by an IRS controller. 

The channel coefficients between AP and IRS, AP and user, AP and eavesdropper, IRS and user and IRS and eavesdropper are denoted by $\mathbf{G} \in \mathbb{C}^{N \times M}$, $\mathbf{h}_{\mathrm{au}} \in \mathbb{C}^{1 \times M}$, $\mathbf{h}_{\mathrm{ae}} \in \mathbb{C}^{1 \times M}$, $\mathbf{h}_{\mathrm{iu}} \in \mathbb{C}^{1 \times N}$, $\mathbf{h}_{\mathrm{ie}} \in \mathbb{C}^{1 \times N}$, respectively. Consider that all channels in the system experience quasi-static flat-fading and the global channel state information (CSI) is perfectly known at both AP and the IRS controller. $n_{\mathrm{U}}$ and $n_{\mathrm{E}}$ denote the additive Gaussian noises at the user and the eavesdropper with zero mean and variance $\sigma_{\mathrm{u}}^{2}$ and $\sigma_{\mathrm{e}}^{2}$, respectively. 

The AP transmits confidential message $\mathbf{s}$ with mean value $0$ and unit variance to the user through beamforming. The beamforming vector is denoted by $\mathbf{f} \in \mathbb{C}^{M \times 1}$ satisfying the constraint
\begin{equation}
\|\mathbf{f}\|^{2} \leq P_{\mathrm{t}},
\end{equation}
where $P_{\mathrm{t}}$ is the maximum transmit power budget at AP.

The vector of the reflection coefficients by the IRS units is denoted by $\boldsymbol{\varphi} \triangleq[\varphi_1, \varphi_2,\dots, \varphi_N]^{T}$, where $\varphi_{n}=\beta_{n} e^{j \theta_{n}}$. $\beta_{n}$ and $\theta_{n}$ stand for the amplitude and the phase shift of the $n$th reflection coefficient, respectively. For  simplicity, we assume ideal hardware configuration for the IRS, meaning that the elements are designed for maximum reflection \cite{cui2019secure}, i.e., $\beta_{n}=1$ and $\theta_{n} \in[0,2 \pi)$, for $n=1,\dots,N$.

The achievable rates at the user and the eavesdropper are, respectively, given by \cite{khisti2008secure}
\begin{align}
R_{\mathrm{u}}=\log _{2}\left(1+\frac{\left|\left(\mathbf{h}_{\mathrm{iu}} \mathbf{\Phi} \mathbf{G}+\mathbf{h}_{\mathrm{au}}\right) \mathbf{f}\right|^{2}}{\sigma_{\mathrm{u}}^{2}}\right),\\
R_{\mathrm{e}}=\log _{2}\left(1+\frac{\left|\left(\mathbf{h}_{\mathrm{ie}} \mathbf{\Phi} \mathbf{G}+\mathbf{h}_{\mathrm{ae}}\right) \mathbf{f}\right|^{2}}{\sigma_{\mathrm{e}}^{2}}\right),
\end{align}
where $\mathbf{\Phi} \triangleq \operatorname {diag}({\boldsymbol\varphi})$. With $R_\mathrm{u}$ and $R_\mathrm{e}$, the secrecy rate from AP to the user is given by \cite{khisti2008secure} 
\begin{equation}\label{u-e}
R_{\mathrm{sec}}=\left[R_{\mathrm{u}}-R_{\mathrm{e}}\right]^{+}.
\end{equation}
The operator $[ \cdot ]^+$ will be omitted in the following expressions since the optimal secrecy rate must be nonnegative.

Our objective is to find the optimal phase-shifts of reflectors in the IRS and corresponding beamforming vector for maximizing the secrecy rate \eqref{u-e}. Thus, the related optimization problem is formulated as
\begin{subequations}\label{opti_prob}
\begin{align}
\begin{split}\label{opti}
    \max _{\mathbf{f}, {\boldsymbol\varphi}} \quad &\log _{2}\left(1+\frac{\left|\left(\mathbf{h}_{\mathrm{iu}} \mathbf{\Phi} \mathbf{G}+\mathbf{h}_{\mathrm{au}}\right) \mathbf{f}\right|^{2}}{\sigma_{\mathrm{u}}^{2}}\right)\\
    \quad&-\log _{2}\left(1+\frac{\left|\left(\mathbf{h}_{\mathrm{ie}} \mathbf{\Phi} \mathbf{G}+\mathbf{h}_{\mathrm{ae}}\right) \mathbf{f}\right|^{2}}{\sigma_{\mathrm{e}}^{2}}\right),
\end{split}\\
  \text{s.t.}\quad& \|\mathbf{f}\|^{2} \leq P_{\mathrm{t}},\label{flimit}\\
   &\left|\varphi_{n}\right|=1, \forall n \label{philimit}
\end{align}
\end{subequations}
Constraint \eqref{flimit} limits the transmission power of the beamforming vector, whereas \eqref{philimit} guarantees maximum reflection. The objective (\ref{opti}) is a non-convex function with regard to $\mathbf{f}$ and $\boldsymbol\varphi$. It is worth noting that the global optimal solution of non-convex optimization problems with unit modulus constraints is usually hard to find. Therefore, \cite{cui2019secure} proposed an iterative optimization method. 

\section{Proposed Solution}\label{sec_sol}
In this section, we present both the traditional convex optimization based algorithm based on \cite{cui2019secure} and the proposed deep learning method to find the optimal phase shifts of IRS units as well as the transmit beamforming vector. 

\setcounter{equation}{20}
\begin{figure*}[b]
\hrule
\begin{equation}
f\left({\boldsymbol\varphi}\right) \triangleq \frac{\frac{1}{\sigma_{u}^{2}}\left({\boldsymbol\varphi}^{H}\mathbf{K}_{\mathrm{u}}^{*}\mathbf{F}\mathbf{K}_{\mathrm{u}}^{T}{\boldsymbol\varphi}
    +\mathbf{h}_{\mathrm{au}}^{*}\mathbf{F}\mathbf{K}_{\mathrm{u}}^{T}{\boldsymbol\varphi}
    +{\boldsymbol\varphi}^{H}\mathbf{K}_{\mathrm{u}}^{*}\mathbf{F}\mathbf{h}_{\mathrm{au}}^{T}
    +\mathbf{h}_{\mathrm{au}}^{*}\mathbf{F}\mathbf{h}_{\mathrm{au}}^{T}\right)
    +1}{\frac{1}{\sigma_{\mathrm{e}}^{2}}\left({\boldsymbol\varphi}^{H}\mathbf{K}_{\mathrm{e}}^{*}\mathbf{F}\mathbf{K}_{\mathrm{e}}^{T}{\boldsymbol\varphi}
    +\mathbf{h}_{\mathrm{ae}}^{*}\mathbf{F}\mathbf{K}_{\mathrm{e}}^{T}{\boldsymbol\varphi}
    +{\boldsymbol\varphi}^{H}\mathbf{K}_{\mathrm{e}}^{*}\mathbf{F}\mathbf{h}_{\mathrm{ae}}^{T}
    +\mathbf{h}_{\mathrm{ae}}^{*}\mathbf{F}\mathbf{h}_{\mathrm{ae}}^{T}\right )
     +1} \label{phiopti_obj}
\end{equation}
\end{figure*}
\setcounter{equation}{8}

\subsection{Conventional Approach}\label{ssec_conventional}
The basic process of alternating optimization is to find the optimal solution for one variable while keeping the others fixed. In this case, we firstly optimize $\mathbf{f}$ assuming that $\boldsymbol\varphi$ is given and then optimize $\boldsymbol\varphi$ with given $\mathbf{f}$ iteratively \cite{cui2019secure}. 

\subsubsection{Optimizing $\mathbf{f}$ with given $\boldsymbol\varphi$}
The optimization problem \eqref{opti_prob} with given $\boldsymbol\varphi$ reduces to
\begin{subequations}\label{opti_prob_f}
\begin{align}\label{4.3}
    & \max _{\mathbf{f}}\quad \frac{\frac{1}{\sigma_{u}^{2}}\left|\left(\mathbf{h}_{\mathrm{iu}} \mathbf{\Phi} \mathbf{G}+\mathbf{h}_{\mathrm{au}}\right) \mathbf{f}\right|^{2}+1}{\frac{1}{\sigma_{\mathrm{e}}^{2}}\left|\left(\mathbf{h}_{\mathrm{ie}} \mathbf{\Phi} \mathbf{G}+\mathbf{h}_{\mathrm{ae}}\right) \mathbf{f}\right|^{2}+1}\\
    & \quad{\rm s.t.}\quad  \|\mathbf{f}\|^{2} \leq P_{\mathrm{t}}.
\end{align}
\end{subequations}
Introducing matrix variables {\bf A} and {\bf B} defined as
\begin{align}
&\mathbf{A}=\frac{1}{\sigma_{\mathrm{u}}^{2}}\left(\mathbf{h}_{\mathrm{iu}} \mathbf{\Phi} \mathbf{G}+\mathbf{h}_{\mathrm{au}}\right)^{H}\left(\mathbf{h}_{\mathrm{iu}} \mathbf{\Phi} \mathbf{G}+\mathbf{h}_{\mathrm{au}}\right), \\
&\mathbf{B}=\frac{1}{\sigma_{\mathrm{e}}^{2}}\left(\mathbf{h}_{\mathrm{ie}} \mathbf{\Phi} \mathbf{G}+\mathbf{h}_{\mathrm{ae}}\right)^{H}\left(\mathbf{h}_{\mathrm{ie}} \mathbf{\Phi} \mathbf{G}+\mathbf{h}_{\mathrm{ae}}\right),
\end{align}
the optimization problem \eqref{opti_prob_f} can be rewritten as
\begin{subequations}
\begin{align}
    &\max_{\mathbf{f}}\quad
    \frac{\mathbf{f}^{H}\mathbf{A}\mathbf{f}+1}
    {\mathbf{f}^{H}\mathbf{B}\mathbf{f}+1}\\
    &\quad{\rm s.t.}\quad {\mathbf{f}^{H}}\mathbf{f}\leq{P_{t}}.
\end{align}
\end{subequations}
Assuming that $\mathbf{e}_{\max}$ is the normalized eigenvector corresponding to the maximum eigenvalue of matrix $\mathbf{C}$ defined as
\begin{equation}
    \mathbf{C}=\left(\mathbf{B}+\frac{1}{P_{t}}\mathbf{I}\right)^{-1}\left(\mathbf{A}+\frac{1}{P_{t}}\mathbf{I}\right),
\end{equation}
where $\mathbf{I}$ is an identity matrix, the optimal solution for $\mathbf{f}$ is given by \cite{khisti2008secure}
\begin{equation}
\label{fopt}
    \mathbf{f}_{\text{opt}}=\sqrt{P_{t}}\mathbf{e}_{\max}.
\end{equation}

\subsubsection{Optimizing $\boldsymbol{\varphi}$ with given $\mathbf{f}$}
\label{phi}
From (\ref{opti_prob}), the optimization problem with given $\mathbf{f}$ can be expressed as:
\begin{subequations}\label{prob_phi}
\begin{align}\label{4.9}
    & \max _{{\boldsymbol\varphi}}\quad \frac{\frac{1}{\sigma_{u}^{2}}\left|\left(\mathbf{h}_{\mathrm{iu}} \mathbf{\Phi} \mathbf{G}+\mathbf{h}_{\mathrm{au}}\right) \mathbf{f}\right|^{2}+1}{\frac{1}{\sigma_{\mathrm{e}}^{2}}\left|\left(\mathbf{h}_{\mathrm{ie}} \mathbf{\Phi} \mathbf{G}+\mathbf{h}_{\mathrm{ae}}\right) \mathbf{f}\right|^{2}+1}\\
    & \quad{\rm s.t.}\quad  \left|\varphi_{n}\right|=1, \forall n.
\end{align}
\end{subequations}
It is known that
\begin{align}
&\mathbf{h}_{\mathrm{iu}} \mathbf{\Phi} \mathbf{G}={\boldsymbol\varphi}^{T} \operatorname{diag}\left(\mathbf{h}_{\mathrm{iu}}\right) \mathbf{G},\\
&\mathbf{h}_{\mathrm{ie}} \mathbf{\Phi} \mathbf{G}={\boldsymbol\varphi}^{T} \operatorname{diag}\left(\mathbf{h}_{\mathrm{ie}}\right) \mathbf{G}.
\end{align}
Let $\operatorname{diag}\left(\mathbf{h}_{\mathrm{iu}}\right)\mathbf{G}=\mathbf{K}_{\mathrm{u}}$, and $\operatorname{diag}\left(\mathbf{h}_{\mathrm{ie}}\right)\mathbf{G}=\mathbf{K}_{\mathrm{e}}$, equations above can be rewritten as
\begin{align}
&\mathbf{h}_{\mathrm{iu}} \mathbf{\Phi} \mathbf{G}={\boldsymbol\varphi}^{T}\mathbf{K}_{\mathrm{u}},\\
&\mathbf{h}_{\mathrm{ie}} \mathbf{\Phi} \mathbf{G}={\boldsymbol\varphi}^{T}\mathbf{K}_{\mathrm{e}},
\end{align}
Defining $\mathbf{f}^{*} \mathbf{f}^{T} \triangleq \mathbf{F}$, problem \eqref{prob_phi} is equivalent to
\begin{subequations}\label{phiopti}
\begin{align}
    \max _{{\boldsymbol\varphi}}  ~&~ f\left({\boldsymbol\varphi}\right) \\
    \quad{\rm s.t.}  ~&~  \left|\varphi_{n}\right|=1, \forall n,
\end{align}
\end{subequations}
where  $f\left({\boldsymbol\varphi}\right)$ is defined in \eqref{phiopti_obj} (at the bottom of the page). 
Let us now define the variables $\mathbf{\nu}_{\mathrm{U}}$, $\mathbf{\nu}_{\mathrm{E}}$, $\mathbf{\Gamma}_{\mathrm{U}}$ and $\mathbf{\Gamma}_{\mathrm{E}}$ as
\setcounter{equation}{21}
\begin{gather}
    \mathbf{\nu}_{\mathrm{U}}=\frac{\mathbf{h}_{\mathrm{au}}^{T} \mathbf{F} \mathbf{h}_{\mathrm{au}}^{*}}{\sigma_{\mathrm{u}}^{2}},
    \quad \mathbf{\nu}_{\mathrm{E}}=\frac{\mathbf{h}_{\mathrm{ae}}^{T} \mathbf{F}     \mathbf{h}_{\mathrm{ae}}^{*}}{\sigma_{\mathrm{e}}^{2}},\label{nu}\\
    \mathbf{\Gamma}_{\rm U} = \frac{1}{\sigma_{u}^{2}}\left[\begin{array}{cc}
\mathbf{K}_{\mathrm{u}}^{T}\mathbf{F}\mathbf{K}_{\mathrm{u}}^{*} & \mathbf{K}_{\mathrm{u}}^{T}\mathbf{F}\mathbf{h}_{\mathrm{au}}^{*} \\
\mathbf{h}_{\mathrm{au}}^{T}\mathbf{F}\mathbf{K}_{\mathrm{u}}^{*} & 0
\end{array}\right],\\
\mathbf{\Gamma}_{\rm E}= \frac{1}{\sigma_{e}^{2}}\left[\begin{array}{cc}
\mathbf{K}_{\mathrm{e}}^{T}\mathbf{F}\mathbf{K}_{\mathrm{e}}^{T} & \mathbf{K}_{\mathrm{e}}^{T}\mathbf{F}\mathbf{h}_{\mathrm{ae}}^{*} \\
\mathbf{h}_{\mathrm{ae}}^{T}\mathbf{F}\mathbf{K}_{\mathrm{e}}^{*} & 0\label{tao}
\end{array}\right].
\end{gather}
By substituting (\ref{nu}) - (\ref{tao}), (\ref{phiopti}) can be simplified as the following equivalent problem
\begin{subequations}\label{4.19}
\begin{align}
    &\max _{\mathbf{v}}\quad \frac{\mathbf{v}^{H} \mathbf{\Gamma}_{\mathrm{U}} \mathbf{v}+\mathbf{\nu}_{\mathrm{U}}+1}{\mathbf{v}^{H} \mathbf{\Gamma}_{\mathrm{E}} \mathbf{v}+\mathbf{\nu}_{\mathrm{E}}+1},\\
    &\quad{\rm s.t.}\quad \mathbf{v}^{H} \mathbf{U}_{n} \mathbf{v}=1, \forall n,
\end{align}
\end{subequations}
where $\mathbf{v}=\left[{\boldsymbol\varphi}^{T}, 1\right]^{T}$ and $\mathbf{U}$ is a three dimensional matrix, with elements of $\mathbf{U}_{n}$ given by
\begin{equation}
    \left[\mathbf{U}_{n}\right]_{i, j}=\left\{\begin{array}{ll}
1 & i=j=n \\
0 & \text { otherwise }.
\end{array}\right.
\end{equation}
Note that problem (\ref{4.19}) is still non-convex. Defining $\mathbf{V} \triangleq \mathbf{\mathbf{v} \mathbf{v}}^{H}$ requires that $\operatorname{rank}(\mathbf{V})\leq 1$. Ignoring this rank constraint, the semidefinite relaxation technique can be applied to address the non-convex problem \cite{jrnl_swipt}. Thus the optimization problem can be reformulated as
\begin{align} 
    &\max_{\mathbf{V} \succeq \mathbf{0} }\quad \frac{\operatorname{tr}\left(\mathbf{\Gamma}_{\mathrm{U}} \mathbf{V}\right)+\mathbf{\nu}_{\mathrm{U}}+1}{\operatorname{tr}\left(\mathbf{\Gamma}_{\mathrm{E}} \mathbf{V}\right)+\mathbf{\nu}_{\mathrm{E}}+1} \\  &\quad{\rm s.t.}\quad\left.\operatorname{tr}\mathbf{( U}_{n} \mathbf{V}\right)=1, \forall n.
\end{align}
Then, we can apply Charnes-Cooper transformation \cite{charnes1962programming} to transform it into a convex semidefinite programming (SDP) problem by defining $\mu=1 /\left[\operatorname{tr}\left(\mathbf{\Gamma}_{\mathrm{E}} \mathbf{V}\right)+\mathbf{\nu}_{\mathrm{E}}+1\right]$. The equivalent optimization problem is given by
\begin{align}
    \max _{\mu \geq 0, \mathbf{Z} \succeq 0} \quad &\operatorname{tr}\left(\mathbf{\Gamma}_{\mathrm{U}} \mathbf{Z}\right)+\mu\left(\mathbf{\nu}_{\mathrm{U}}+1\right) \\
    \quad\text { s.t. } \quad &\operatorname{tr}\left(\mathbf{\Gamma}_{\mathrm{E}} \mathbf{Z}\right)+\mu\left(\mathbf{\nu}_{\mathrm{E}}+1\right)=1 \\
     &\operatorname{tr}\left(\mathbf{U}_{n} \mathbf{Z}\right)=\mu, \forall n,
\end{align}
where $\mathbf{Z}=\mu\mathbf{V}$. The problem is now convex and can be efficiently solved by interior-point methods (e.g., CVX) \cite{polik2010interior}. Gaussian randomization method can be used to cope with the rank constraint and obtain an approximate optimal solution.

\subsection{Proposed Deep Learning Algorithm}\label{ssec_dnn}
While deep learning has played an important role in many applications in recent years, there are a number of challenges facing the design of a DL method for the secure communication scenario under consideration \cite{jrnl_irs_sec}:

\begin{itemize}
    \item It is difficult to model the input-output relationship of DNN due to the large number of parameters involved in the calculation of secrecy rate. 
    \item Acquisition of training data set, in particular, supervised learning not only needs a lot of channel samples, but also the target outputs mapped with those samples. 
    \item Excepting the transmit power and noise power, all the other parameters are inherently complex; nevertheless, most of the deep neural network technologies are based on real-valued operation and representation.
\end{itemize}

Deep neural network (DNN) is a kind of mathematical framework that realizes the mapping from input to output through a series of data transformation layers. Supervised learning and unsupervised learning are widely used  algorithms in machine learning. In this paper, we use supervised learning to train the DNN.

\begin{figure}[ht]
\centering
\includegraphics[width=\linewidth]{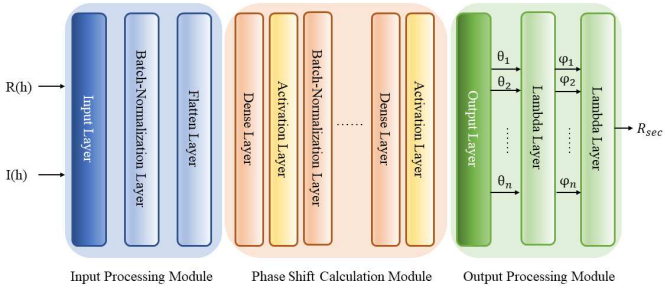}
\caption{Proposed DNN framework.}\label{DNNmodel}
\end{figure}

The proposed DNN framework used to obtain the optimal phase shifts of IRS reflection units is shown in Fig.~\ref{DNNmodel}, and the beamforming vector $\mathbf{f}$ is computed according to (\ref{fopt}). For handling multiple inputs, we use Keras functional API model in this work because of its flexibility in dealing with multi-inputs problem. The proposed neural network includes three modules: input processing module, phase shift calculation module and output processing module.  

The input processing module has several parallel sets of input lines to take channel coefficients as inputs. The complex channel coefficients are divided into real part and imaginary part since the neural network framework can only perform real-valued operations. After each set of inputs passing through batch normalization layer and flatten layer in turn, they will be concatenated into one set of one-dimensional data that is taken as the input to the phase shift calculation module. In the input processing module, each layer is composed of a fixed number of neurons which is determined by the number of channel coefficients.

The phase shift calculation module includes several dense layers and batch normalization layers. The number of neurons in each dense layer is adjustable. The function of all these dense layers is to establish the logical relationship between the channel coefficients and the phase shifts of IRS units.

The output processing module is composed of one dense layer and two Lambda layers. The dense layer outputs the phase shifts of the IRS units in radians. In the first Lambda layer, the Euler formula is used to convert the phase shift $\theta_n$ into complex form $\varphi_n$, which also satisfies the constrains defined in (\ref{philimit}), $|\varphi_{n}|=1$; the second Lambda layer takes all the channel coefficients, transmit power and ${\boldsymbol\varphi}$ as inputs to obtain the secrecy rate.

\subsubsection*{Learning Policy}\label{policy}
\begin{figure}[ht]
\centering
\includegraphics[width=0.8\linewidth]{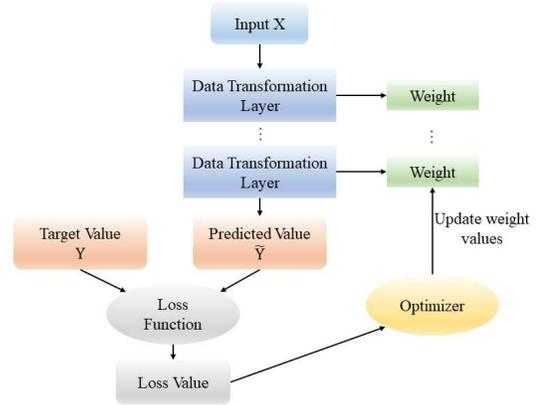}
\protect\caption{Supervised learning algorithm.} \label{DNNsupervised}
\end{figure}

Fig.~\ref{DNNsupervised} shows how DNN based on supervised learning works. The input $X$ will go through several data transformation layers, and the predicted output $\tilde{Y}$ will be generated. The loss function is generated by comparing the output value $\tilde{Y}$ and actual target value $Y$. Then, the optimizer will iteratively optimize the weight values in each layer based on the loss value. For supervised learning, there are two dense layers and one batch-normalization layer. We take Adam as the optimizer and choose mean squared error (MSE) as the loss function, given by
\begin{equation}
    MSE=\frac{1}{n} \sum_{i=1}^{n}\left(\tilde{y}_{i}-t_{i}\right)^{2}
\end{equation}
where $\tilde{y_i}$ is the output of the neural network, $t_i$ is the corresponding training targets, $i$ is the index of data. The channel coefficients and target values are generated by computer simulation using the alternating optimization algorithm in Section~\ref{ssec_conventional}.

\section{Numerical Simulations}\label{sec_sim}
In this section, we perform numerical simulations to demonstrate the effectiveness of the proposed supervised learning based secrecy rate maximization approach for the proposed scenario of interest. We first demonstrate the learning accuracy of the proposed DL method to determine the optimal hyper parameters for the proposed wiretapping scenario. We then compare the performance of the trained DNN against the alternating optimization algorithm in \cite{cui2019secure} and two other benchmark schemes as defined below:
\begin{itemize}
    \item \textbf{Optimal AP without IRS}: The optimal beamforming vector $\mathbf{f}$ is computed according to \eqref{fopt} with ${\boldsymbol\varphi}=0$.
    \item \textbf{AP MEV with IRS}: Firstly, we set beamforming vector $\mathbf{f}$ same as the ‘Optimal AP without IRS’ scheme, then obtain optimal ${\boldsymbol\varphi}$ by using the method in section~\ref{phi}.
    \item \textbf{Alternating Optimization}: Based on the method in Section~\ref{ssec_conventional}.
    \item \textbf{Supervised Learning}: Get training data from the simulation results of alternating optimization. The trained DNN can output reflection phase shifts by taking channel coefficients and transmit power as inputs.
\end{itemize}

\begin{figure*}[ht]
\begin{minipage}{0.33\textwidth}
\includegraphics[width=\textwidth]{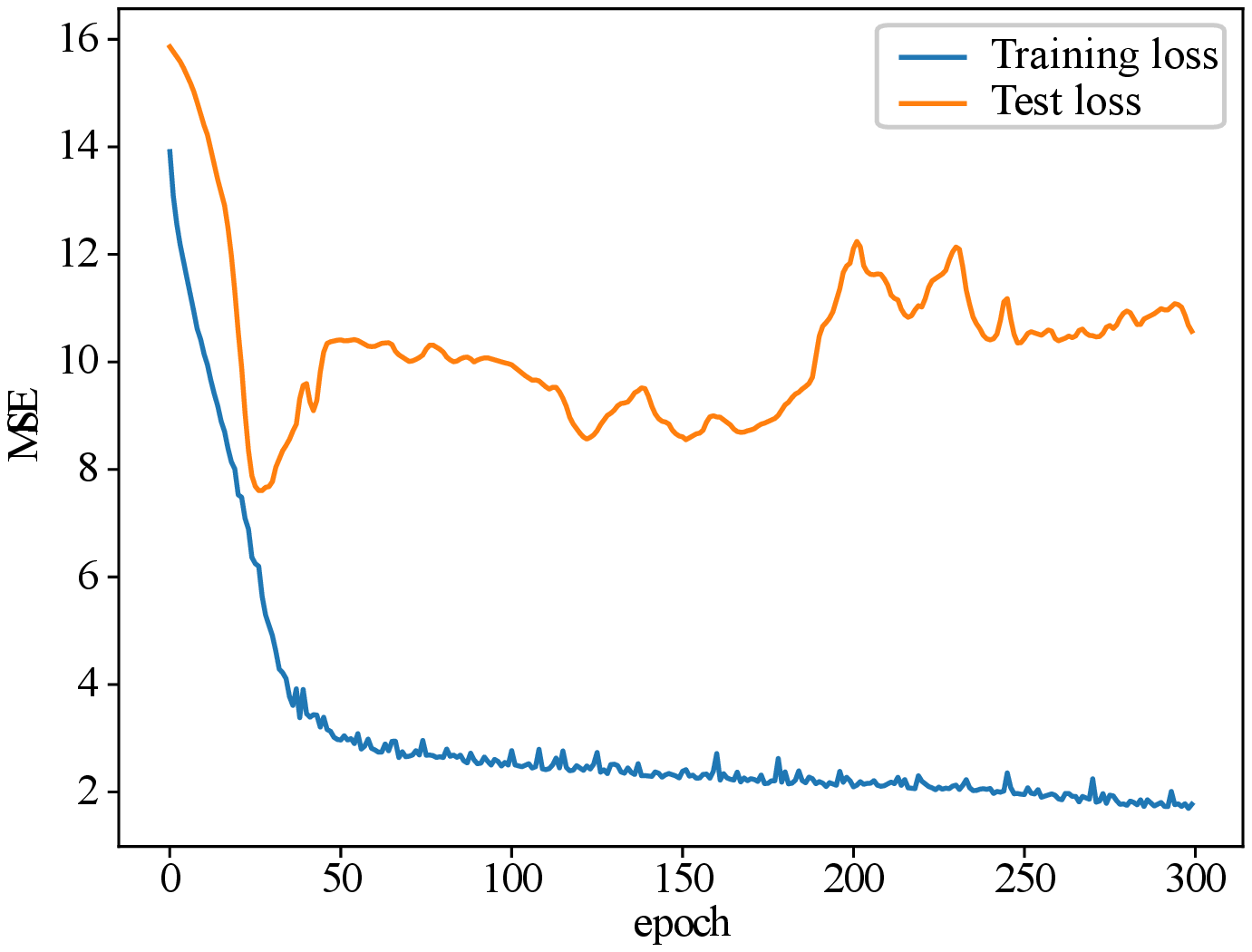}
\caption{Model loss with 100 training sets.}
\label{4-100}
\end{minipage}%
\hfill
\begin{minipage}{0.33\textwidth}
\includegraphics[width=\textwidth]{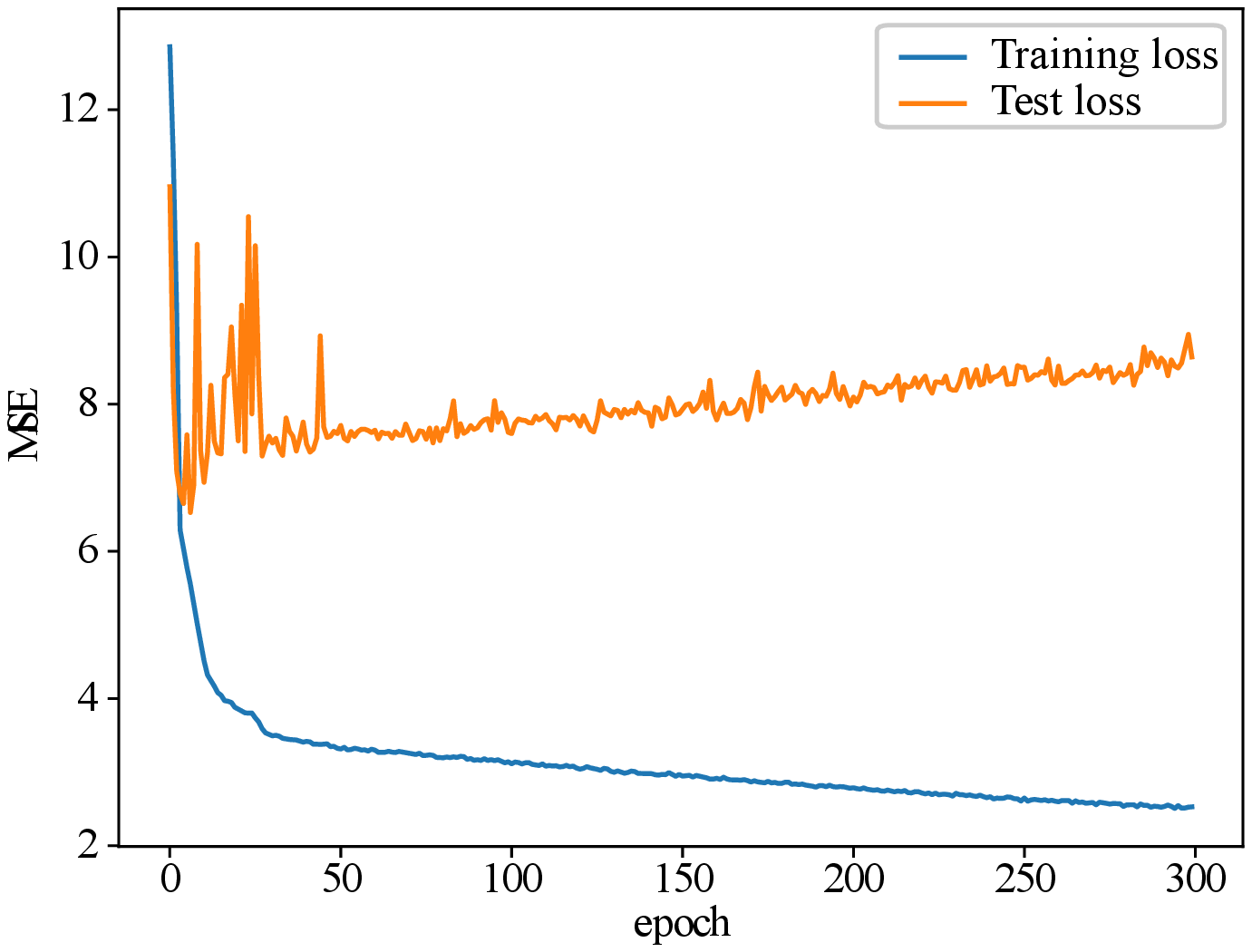}
\caption{Model loss with 2700 training sets.}
\label{4-2700}
\end{minipage}%
\hfill
\begin{minipage}{0.33\textwidth}
\includegraphics[width=\textwidth]{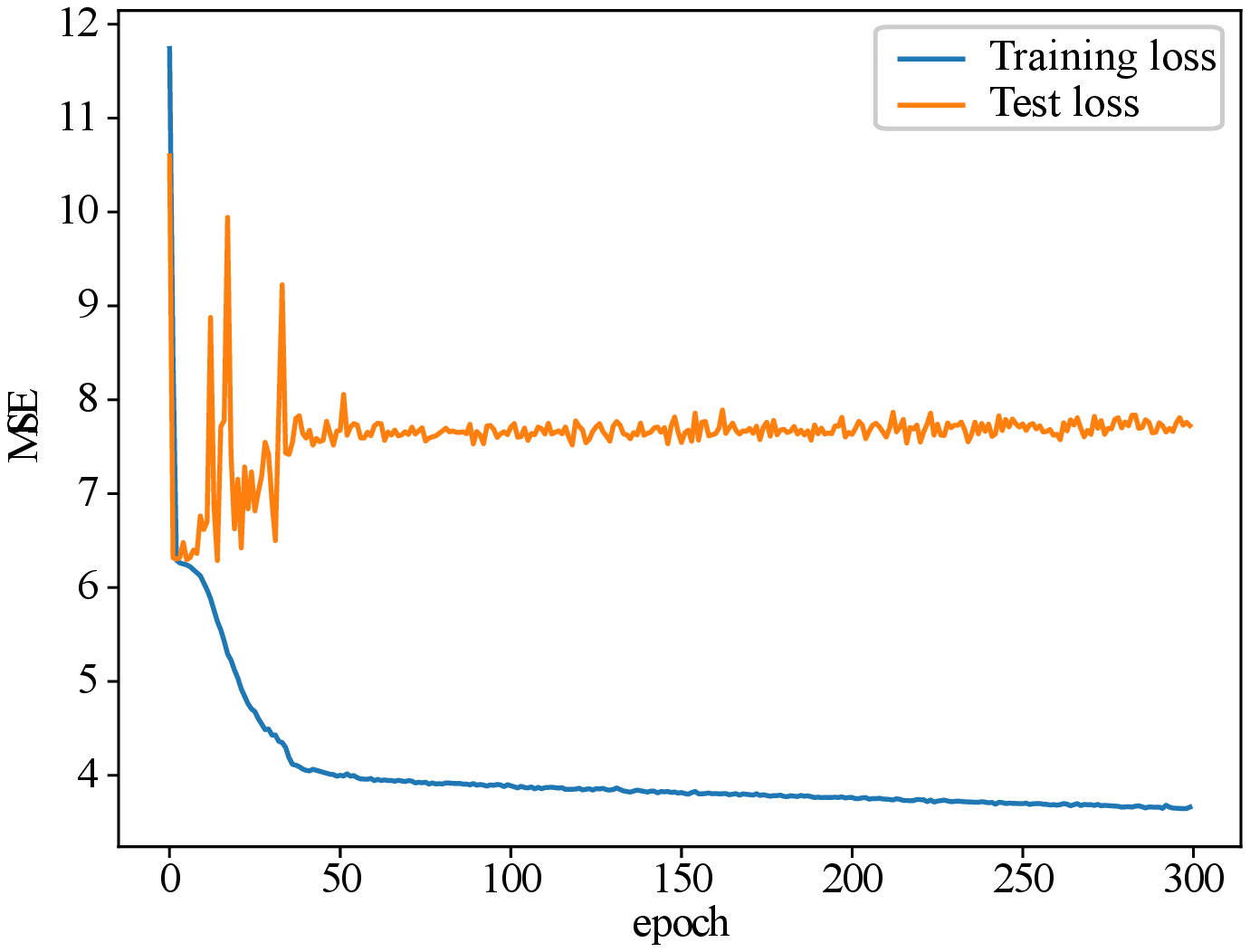}
\caption{Model loss with 9000 training sets.}
\label{4-9000}
\end{minipage}
\end{figure*}

For a fair comparison, we choose parameter settings as in \cite{cui2019secure} wherever applicable. Accordingly, we set $M=4$ and $N=25$, the noise variance at both the eavesdropper and the user are set to $\sigma_{\mathrm{u}}^{2}=\sigma_{\mathrm{e}}^{2}= -80$dBm, unless otherwise specified. The AP, eavesdropper and the user are located on the same horizontal line. The distance between AP and eavesdropper, AP and the user and eavesdropper and the user are denoted by $d_\mathrm{ae}$, $d_\mathrm{au}$ and $d_\mathrm{eu}$, respectively, and set as $d_\mathrm{ae}=145$m and $d_\mathrm{au}=150$m, thus $d_\mathrm{eu} = d_\mathrm{au} - d_\mathrm{ae} = 5$m. The IRS-eavesdropper, IRS-user and AP-IRS link distances are set as $d_\mathrm{ie}=5$m, $d_{\mathrm{iu}}=\sqrt{\left({d_{\mathrm{ie}}}^{2}+{d_{\mathrm{eu}}}^{2}\right)}$ and $d_\mathrm{ai}=\sqrt{\left({d_\mathrm{ae}}^2+{d_\mathrm{ie}}^2\right)}$, respectively. 
Since the AP, eavesdropper and user lie on the same horizontal line, the channels from AP to the user $\mathbf{h}_\mathrm{au}$ and to the eavesdropper $\mathbf{h}_\mathrm{ae}$ are assumed to experience spatially correlated Rician fading, with Rician factors ${K}_\mathrm{au}={K}_\mathrm{ae}=1$ and the spatial correlation matrix $\mathbf{R}$, which is given by $[\mathbf{R}]_{i,j}=r$ where $r=0.95$. The channel coefficients $\mathbf{h}_\mathrm{au}$ and $\mathbf{h}_\mathrm{ae}$ can be obtained as $\mathbf{h}_{\mathrm{au}}=\sqrt{{\eta_{0}}\left(d_{0} / d_{\mathrm{au}}\right)^{\psi_{\mathrm{au}}}} \mathbf{g}_{\mathrm{au}}$, and  $\mathbf{h}_{\mathrm{ae}}=\sqrt{{\eta_{0}}\left(d_{0} / d_{\mathrm{ae}}\right)^{\psi_{\mathrm{ae}}}} \mathbf{g}_{\mathrm{ae}}$, where $\eta_{0}=-30$dB is the path loss with reference distance $d_{0}=1$m, and $\psi_\mathrm{au}=\psi_\mathrm{ae}=3$ are the corresponding path loss exponents. The other channels $\mathbf{G}$, $\mathbf{h}_\mathrm{iu}$ and $\mathbf{h}_\mathrm{ie}$ are independent Rician fading with corresponding path loss $\psi_\mathrm{ai}=2.2$ and $\psi_\mathrm{iu}=\psi_\mathrm{ie}=3$. 

\subsection{Simulation Setup for the Deep Learning Model}
For the proposed supervised learning approach, a total of $10,000$ data samples have been generated, of which $90\%$ is used for training and the remaining $10\%$ for testing the performance, unless otherwise specified. The proposed DNN was trained in a GPU server with the following configuration:\\
- Intel Xeon Scalable Silver 4110 8Core 2.1GHz processor,\\
- 128 GB DDR4 2666 MHz ECC registered memory.

\subsection{Overfitting Problem}
Overfitting is a common problem in deep learning. It refers to the state that only the training data can be well-fitted, but not the data that is not included for training. For supervised learning in particular, over-fitting problem is critical, so we first focus on over-fitting phenomenon in supervised learning and its solutions.

\subsection{Methods to Suppress Overfitting}
\subsubsection{Increasing Training Data}
In Fig.~\ref{4-100}, we used 100 training sets and 100 test sets. To further improve the generalization ability of DNN, the number of training sets is increased from 100 to 2700 in Fig.~\ref{4-2700} and it is tested on 300 data sets. From Fig.~\ref{4-2700}, it is observed that the training loss curve is smoother and the test loss is slightly lower compared with Fig.~\ref{4-100}. 

Even though Fig.~\ref{4-2700} has some improvement compared with Fig.~\ref{4-100}, it still has a large gap between training loss and test loss. Thus, the number of training sets is further increased to 9000. After training it on 9000 data sets and testing it on 1000 data sets, the results are illustrated in Fig.~\ref{4-9000}. It can be observed that the value of test loss decreased to below 8 and no longer showed an upward trend.

\subsubsection{Decreasing Hidden Layers}
\begin{figure*}[ht]
\begin{minipage}[t]{0.33\textwidth}
\includegraphics[width=\textwidth]{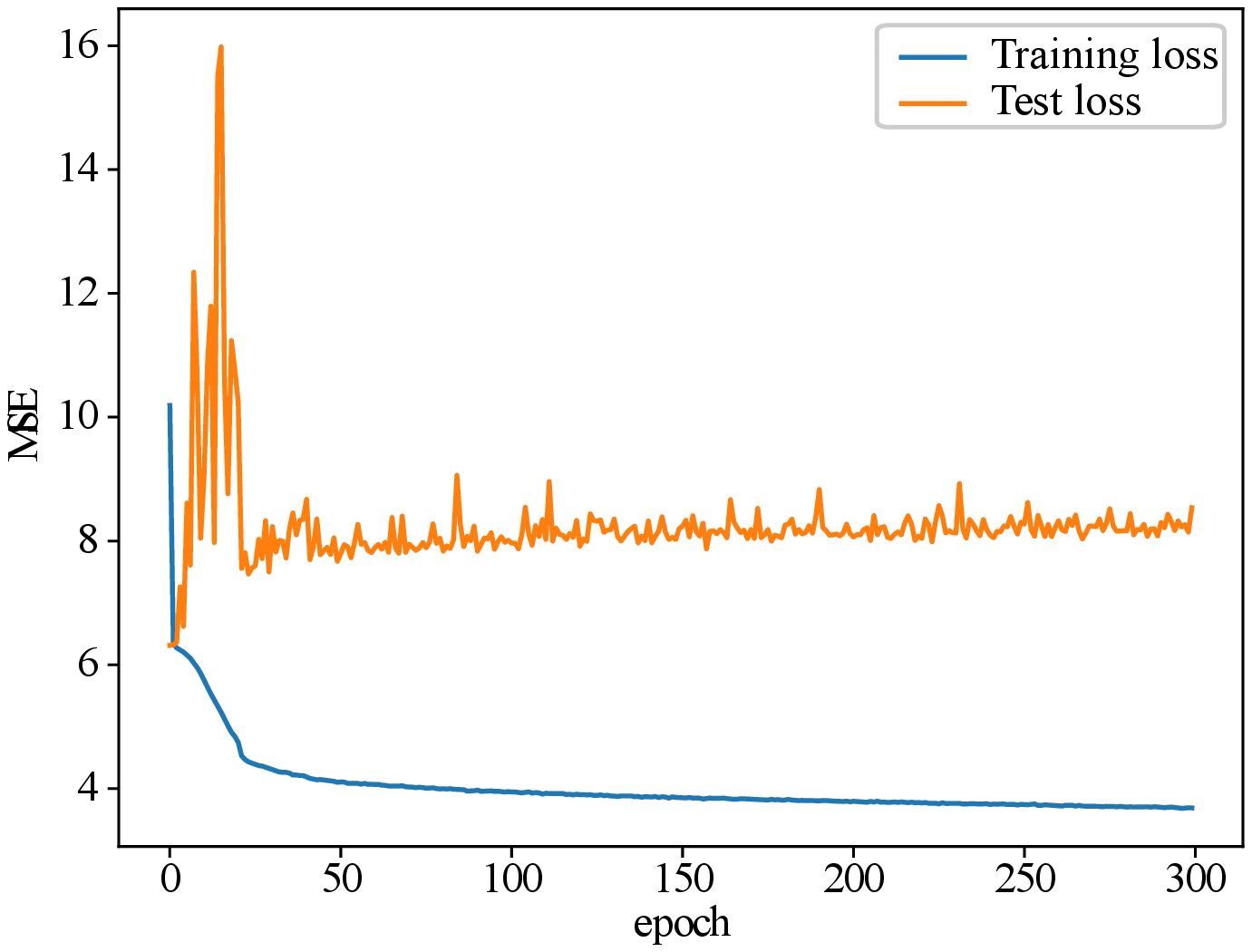}
\caption{Model loss with 3 hidden layers.}
\label{3-9000}
\end{minipage}%
\begin{minipage}[t]{0.33\textwidth}
\includegraphics[width=\textwidth]{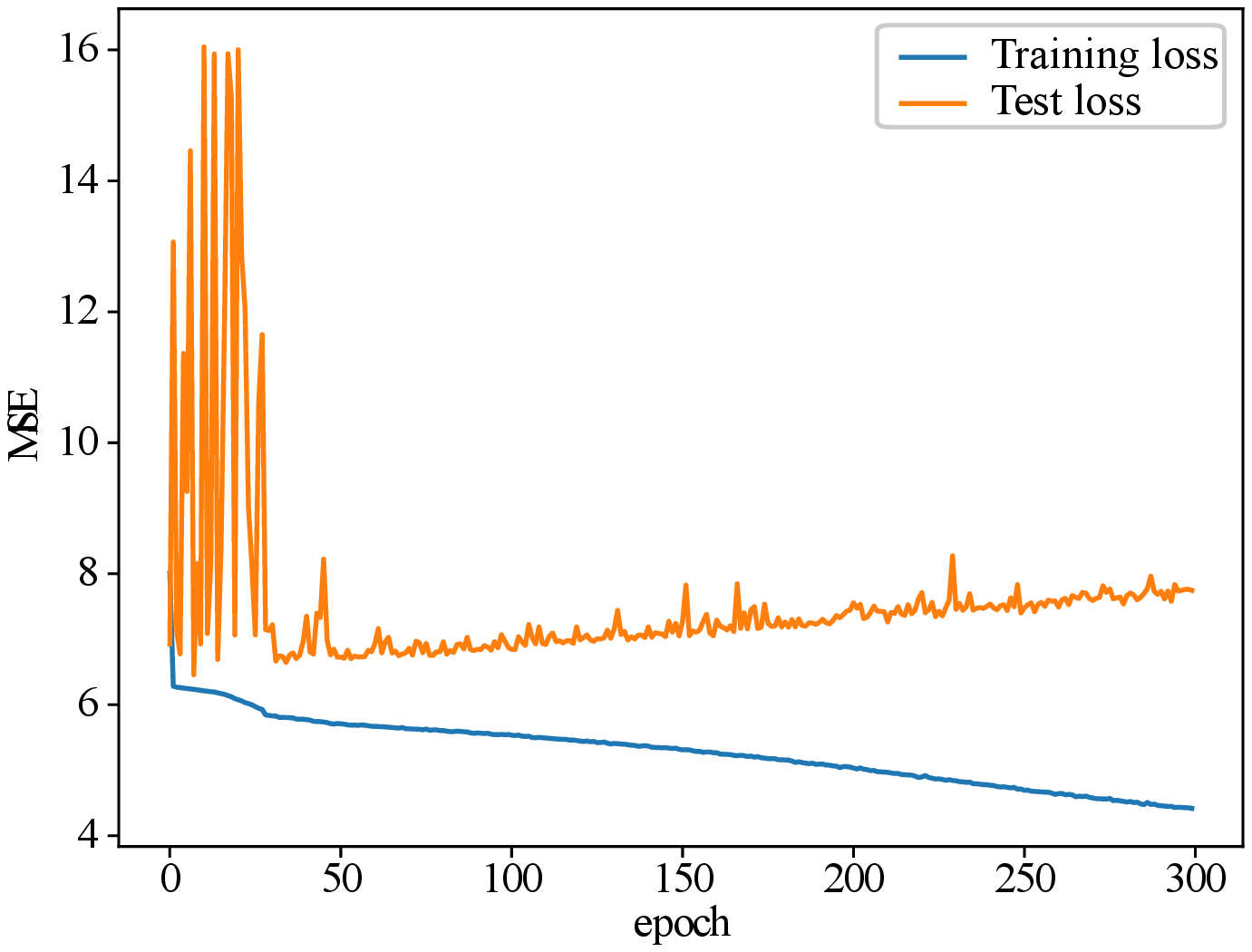}
\caption{Model loss with 2 hidden layers.}
\label{2-9000}
\end{minipage}
\begin{minipage}[t]{0.33\textwidth}
\includegraphics[width=\textwidth]{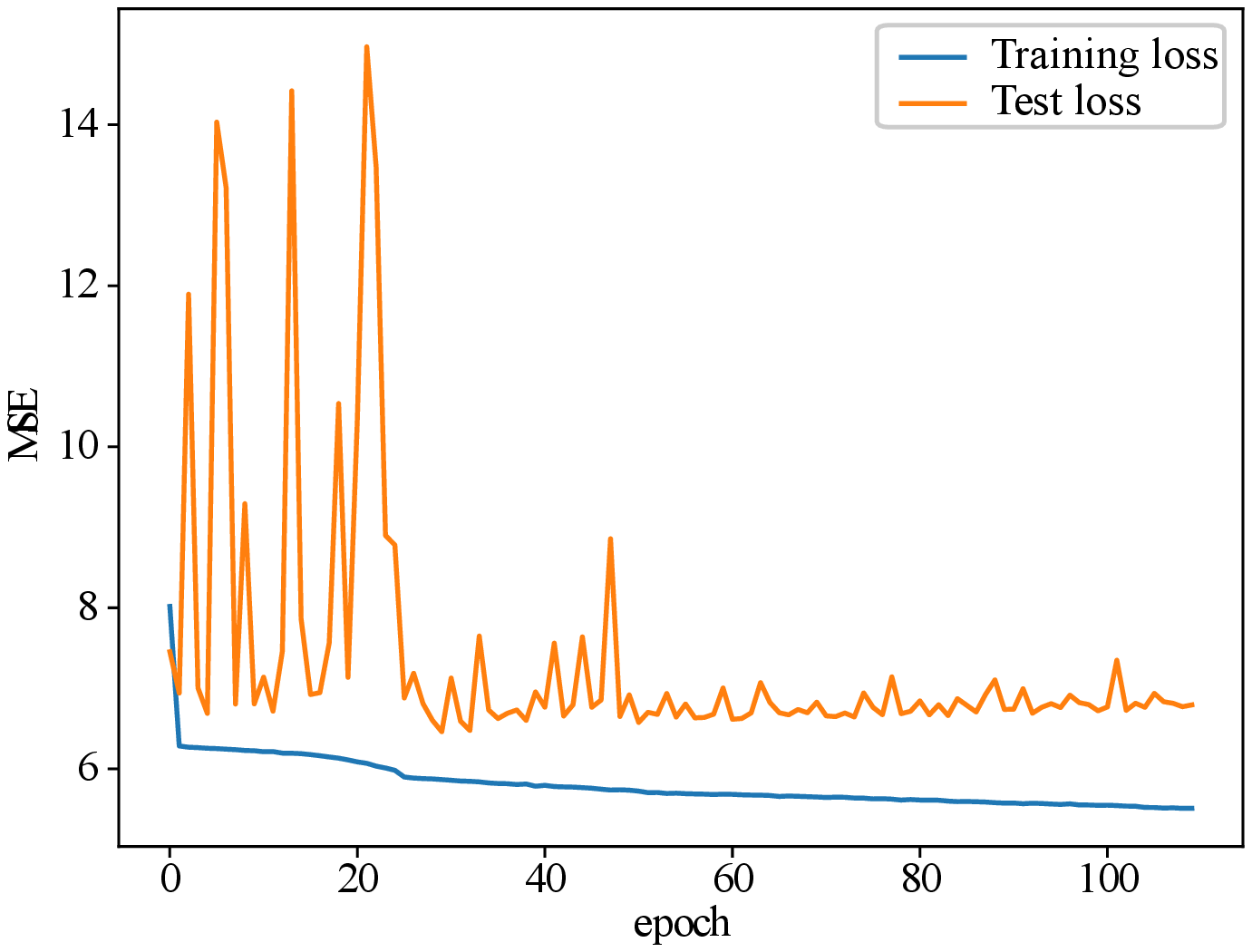}
\protect\caption{Model loss with early stopping.} \label{2-9000-110}
\end{minipage}

\end{figure*}

It is known that the DNN architecture in Fig.~\ref{4-100}, Fig.~\ref{4-2700} and Fig.~\ref{4-9000} has 4 hidden layers. After reducing the number of hidden layers from 4 to 3 and 2, the model loss is shown as Fig.~\ref{3-9000} and Fig.~\ref{2-9000}, respectively. It can be observed from Fig.~\ref{2-9000} that 2 hidden layers greatly reduce the model loss, thereby effectively suppressing overfitting.

\subsubsection{Early Stopping}
It can be observed that the overfitting in Fig.~\ref{2-9000} is slowly  increasing with epoch. However, Fig.~\ref{2-9000-110} illustrates the model loss with early stopping at $110$ epochs. We can observe from Fig.~\ref{2-9000-110} that the test loss is no longer increasing with epochs. 

A comparison between the initial model loss in Fig.~\ref{4-100} and the final model loss in Fig.~\ref{2-9000-110} reveals that overfitting can be effectively suppressed by increasing training data set, decreasing the number of hidden layers stopping early as appropriate.

\begin{figure}[ht]
\centering
\includegraphics[width=0.8\linewidth]{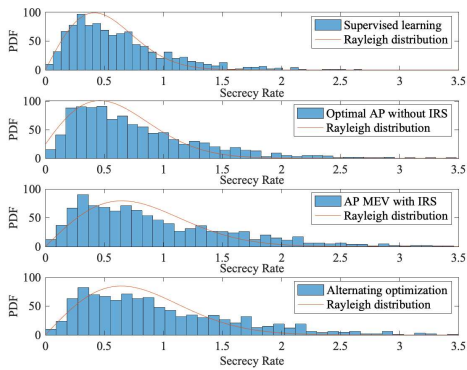}
\caption{PDF of the achievable secrecy rate.}
\label{pdf5}
\end{figure}

\begin{figure}[ht]
\centering
\includegraphics[width=0.8\linewidth]{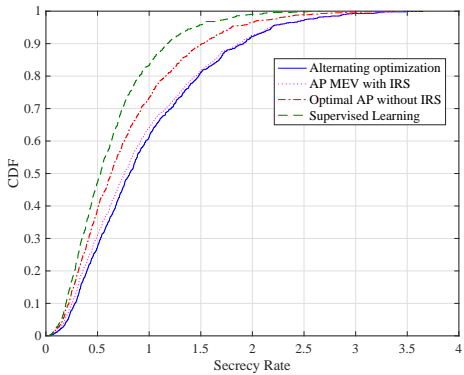}
\protect\caption{CDF of the achievable secrecy rate.} \label{cdf5}
\end{figure}

\subsection{Secrecy Rate Result}
Finally, we illustrate the achievable secrecy rate performance of the conventional and supervised learning approaches. All of the simulation results are averaged over $1000$ channel realizations. 

Fig.~\ref{pdf5} and Fig.~\ref{cdf5} illustrate the PDF and the CDF, respectively, of the secrecy rates obtained by the schemes. The simulation results are obtained from $1,000$ samples with $P_t$ equal to $20$dBm. Again, the alternating optimization yields the best performance and the performance of the 'AP MEV with IRS' is slightly worse than this method. Both the ‘Optimal AP without IRS' and 'Supervised Learning' are also comparable. The marginal performance loss is due to insufficient training data. Theoretically, even if supervised learning can obtain sufficient training data and time, its performance can not surpass the alternating optimization algorithm. This is because the result of the `Alternating optimization' scheme is the benchmark for the supervised learning scheme. Although the conventional optimization algorithms show higher secrecy rates, these schemes are computationally expensive in practice.

\section{Conclusion}\label{sec_con}
The IRS is a cost-effective technology consisting of a large number of low-cost reflection units, which can greatly improve the performance of the physical layer without incurring the high cost and power consumption required for multiple antennas. We have introduced a truly intelligent reflecting surface aided secure communication system. Simulation results demonstrate that the proposed supervised learning approach can achieve comparable performance with the alternating optimization algorithm, and is simpler to implement, shorter in operation time and offers greater flexibility.

\ifCLASSOPTIONcaptionsoff
  \newpage
\fi

\bibliographystyle{IEEEtran}\footnotesize{

\bibliography{IEEEabrv,ref1}}%

\end{document}